\documentclass[final]{aipproc}
\layoutstyle{6x9}
\usepackage{graphicx,amsmath,wrapfig,epsfig}
\usepackage{color}

\begin{document}
\title{Clustering effects for explaining an anomalous JLab result 
       on the $^9$Be structure function}
\author{M. Hirai}{
address={Department of Physics, Faculty of Science and Technology,
Tokyo University of Science \\ 
2641, Yamazaki, Noda, Chiba, 278-8510, Japan}}
\author{S. Kumano}{
  address={KEK Theory Center, Institute of Particle and Nuclear Studies, KEK \\
           and 
           Department of Particle and Nuclear Studies,
           Graduate University for Advanced Studies \\
           1-1, Ooho, Tsukuba, Ibaraki, 305-0801, Japan}}
\author{K. Saito}{
address={Department of Physics, Faculty of Science and Technology,
Tokyo University of Science \\ 
2641, Yamazaki, Noda, Chiba, 278-8510, Japan}}
\author{T. Watanabe}{
address={Department of Physics, Faculty of Science and Technology,
Tokyo University of Science \\ 
2641, Yamazaki, Noda, Chiba, 278-8510, Japan}}

\begin{abstract}
An anomalous nuclear modification was reported by JLab measurements
on the beryllium-9 structure function $F_2$. It is unexpected
in the sense that a nuclear modification slope is too large to be
expected from its average nuclear density. We investigated whether
it is explained by a nuclear clustering configuration
in $^9$Be with two $\alpha$ nuclei and surrounding neutron clouds.
Such clustering aspects are studied by using 
antisymmetrized molecular dynamics (AMD) and also 
by a simple shell model for comparison.
We consider that nuclear structure functions $F_2^A$ consist of 
a mean conventional part and a remaining one depending on 
the maximum local density.
The first mean part does not show a significant cluster effect
on $F_2$. However, we propose that the remaining one could 
explain the anonymous JLab slope, and it is associated with 
high densities created by the cluster formation in $^9$Be.
The JLab measurement is possibly the first signature of clustering 
effects in high-energy nuclear reactions. A responsible physics 
could be an internal nucleon modification, which is caused by 
the high densities due to the cluster configuration.
\end{abstract}

\maketitle

\section{Introduction}

Nuclear modifications of structure functions $F_2$ were found by
the European muon collaboration, and it is often called the EMC effect.
The nuclear modifications are now experimentally measured from
relatively small $x$ to large $x$. Theoretical mechanisms are 
different depending on the $x$ region \cite{sumemc}.
At small $x$, the modifications are caused by nuclear shadowing 
due to multiple scattering of a $q\bar q$ pair coming from 
the virtual photon. At medium $x$, nuclear effects are due to
binding and a possible internal nucleon modification. 
Nucleon Fermi motion and short-range nucleon-nucleon correlation
cause large-$x$ nuclear effects. Using nuclear $F_2$ data
together with the other ones, optimum parton distribution
functions have been determined in nuclei \cite{hkn,nuint09-npdf}.

Nuclear modifications at $x>0.2$ are generally described by 
a convolution model with the nucleon structure function $F_2^N$ 
convoluted with the nucleon momentum distribution in a nucleus. 
The nucleon four-momentum distribution, which is called the spectral 
function, has been calculated in a conventional shell model possibly
with the short-range correlations and internal nucleon
modifications. On the other hand, it is known in
low-energy nuclear physics that some nuclei have
clustering configurations which cannot be described by
simple shell models. It is an an interesting topic to
investigate such clustering aspects in deep-inelastic structure
functions \cite{hksw-cluster}.

\begin{wrapfigure}{r}{0.38\textwidth}
   \vspace{-0.2cm}
   \begin{center}
       \epsfig{file=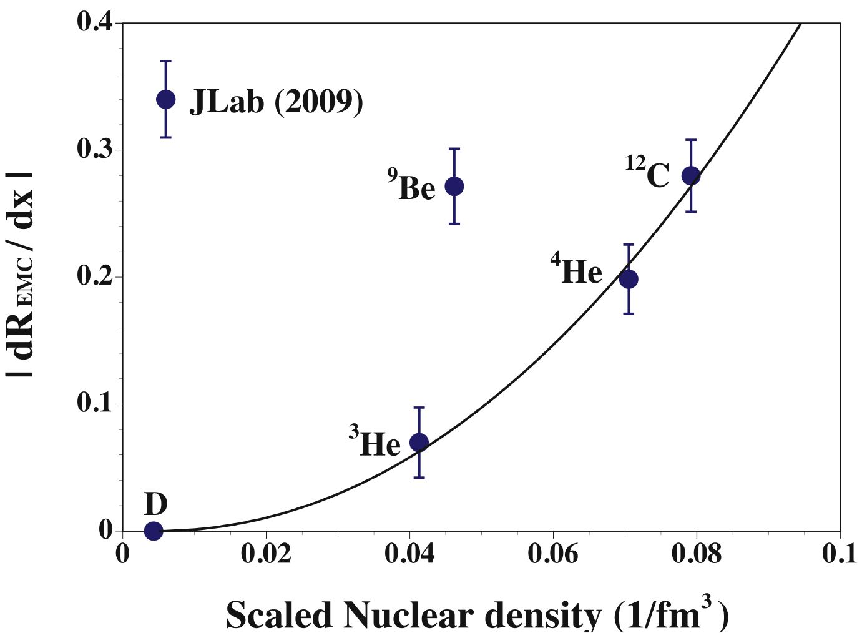,width=0.33\textwidth} \\
   \end{center}
   \vspace{-0.2cm}
{\footnotesize \hspace{0.40cm}
 {\bf FIGURE 1.} Nuclear modification 

\noindent
\hspace{0.40cm} 
  slopes $| d(F_2^A/F_2^D)/dx |$ by JLab.
 }
\vspace{-0.2cm}
\end{wrapfigure}

An anomalous nuclear modification was found for the beryllium-9 nucleus 
at the Thomas Jefferson National Accelerator Facility (JLab)
by measuring ratios $F_2^A/F_2^D$, where $A$ and $D$ denote
a nucleus and the deuteron. Usually, nuclear modifications are 
smooth functions of the average nuclear density. 
JLab measurements showed the nuclear modification 
slope $d(F_2^A/F_2^D)/dx$ by approximating their $F_2^A/F_2^D$
data by straight lines at $0.35<x<0.7$ \cite{jlab-09}. They are shown 
in Fig. 1 as a function of the scaled nuclear density. 
It is obvious from the figure that the nuclear modification slope 
of $^9$Be is too large to be expected from the average density.

In Ref. \cite{hksw-cluster}, we proposed that the anomalous JLab
result could be interpreted by a clustering phenomenon in 
the $^9$Be nucleus. Here, the clustering means that the $^9$Be 
consists of two $\alpha$ ($^4$He) nuclei with extra-neutron clouds.  
Such a nuclear clustering phenomenon had never been investigated 
in the structure functions, although small quark clusters such as
a six-quark bag model was considered in the early stage
of EMC-effect studies. 
We explain that the JLab measurement could be interpreted
if the modification slope is plotted as a function of
the maximum local density at a cluster
by using a theoretical method of antisymmetrized 
molecular dynamics (AMD).
In this article, we explain our theoretical approach
for explaining the anomalous JLab result.

\section{Nuclear clustering effects in $\mathbf F_2$}
\vspace{-0.1cm}

\begin{wrapfigure}{r}{0.43\textwidth}
   \vspace{-0.5cm}
   \begin{center}
       \epsfig{file=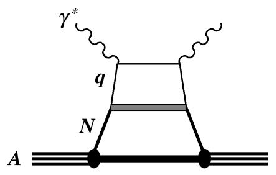,width=0.32\textwidth} \\
   \end{center}
   \vspace{-0.2cm}
{\footnotesize \hspace{0.2cm}
 {\bf FIGURE 2.} Convolution description of $F_2^A$.
 }
\label{fig:convolution}
\vspace{-0.4cm}
\end{wrapfigure}

Nuclear structure functions at $x>0.2$ are described by
a convolution description shown in Fig. 2. The nuclear 
structure function $F_2^A$ is theoretically evaluated
in two steps: first by calculating nucleon ($N$) momentum 
distribution in a nucleus ($A$), second by calculating
the quark ($q$) momentum distribution in the nucleon.
Then, the virtual photon from the charged lepton
(electron in the JLab case) interacts with the quark.
Namely, $F_2^A$ is given by the nucleonic one $F_2^N$ 
convoluted with the nucleon momentum distribution
in the nucleus $f(y)$
\cite{sumemc,ek03}:
\begin{align}
\! \! \! \! \!
F_{2}^A (x, Q^2) & = \int_x^A dy \, f(y) \, F_{2}^N (x/y, Q^2) , \ \ 
f(y)  =  \frac{1}{A} \int d^3 p_N
     \, y \, \delta \left( y - \frac{p_N \cdot q}{M_N \nu} \right) 
     | \phi (\vec p_N) |^2 ,
\label{eqn:w-convolution}
\end{align}
where $y$ is the momentum fraction 
$ y   =  M_A \, p_N \cdot q /(M_N \, p_A \cdot q) $.

In order to illustrate clustering effects on $F_2^A$,
we calculate the momentum density $| \phi (\vec p_N) |^2$
in two methods, a simple shell model with the harmonic-oscillator 
potential $M_N \omega^2 r^2/2$ and the AMD. 
It is known that the wave functions of the harmonic-oscillator 
potential are expressed by Laguerre polynomials
and spherical harmonics.
The AMD (or FMD (fermionic molecular dynamics)) is 
a typical method in investigating clustering aspects 
of nuclear structure.
The advantage of the AMD is that it does not assume 
any specific structure, cluster- or shell-like configuration, 
on nuclei.
The AMD is a variational method, in which
a nuclear wave function is given by the Slater determinant
of single-particle wave packets:
$\left | \Phi (\vec r_1, \vec r_2, \cdot\cdot\cdot, \vec r_A ) \right >
        =  \text{det} [ \varphi_1 (\vec r_1), \varphi_2 (\vec r_2), 
                              \cdot\cdot\cdot, \varphi_A (\vec r_A) ] 
/ \sqrt{A!} $.
Here, $\varphi_i (\vec r_j)$ is the single-particle wave function
given by
$\varphi_i (\vec r_j) = \phi_i (\vec r_j) \, \chi_i \, \tau_i$
with spin and isospin states $\chi_i$ and $\tau_i$. 
The function $\phi_i (\vec r_j)$ is 
the space part expressed as 
$\phi_i (\vec r_j) = ( 2 \nu / \pi )^{3/4}
        \exp  [ - \nu 
           ( \vec r_j - \vec Z_i / \sqrt{\nu} ) ^2  ] $.
Supplying simple $NN$ interactions, we determine
the variational parameters $\nu$ and $\vec Z_i$
by minimizing the system energy.
Calculated AMD space densities are shown in Figs. 3 and 4 
for $^4$He and $^9$Be, respectively. The $^4$He AMD density 
is almost the same as the shell-model one. However,
as shown in Fig. 4, the $^9$Be density is totally different 
from a monotonic shell-model density because the $^9$Be has 
the configuration of two $\alpha$ clusters with surrounding 
neutron clouds.
Using both densities of the shell and AMD models,
we calculate $F_2^A$ in Eq. (\ref{eqn:w-convolution}).

\vspace{0.2cm}
\begin{figure}[h]
   \includegraphics[width=0.36\textwidth]{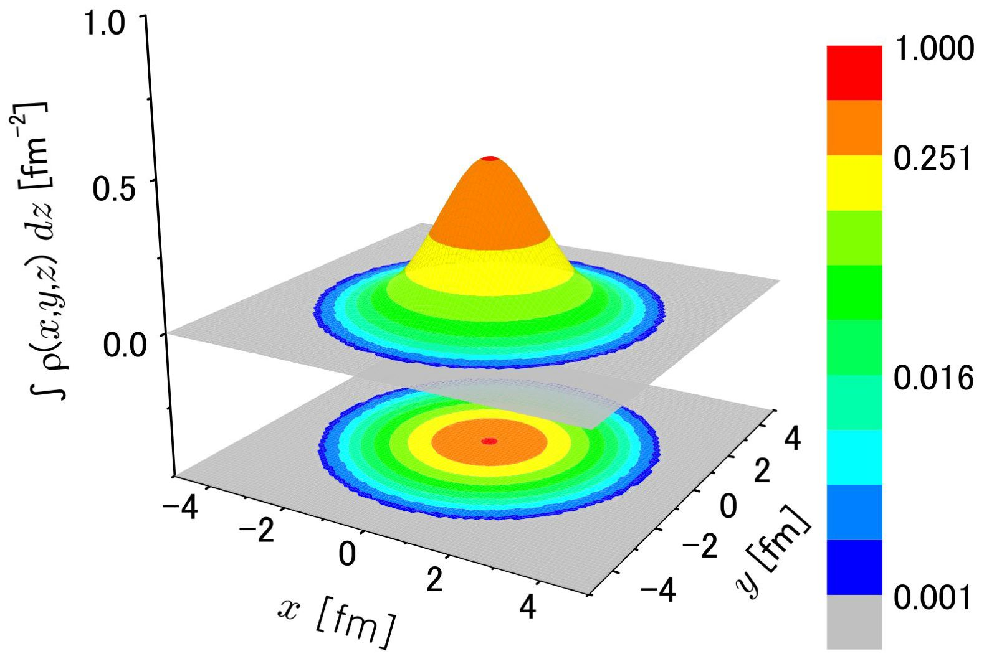}
   \hspace{1.8cm}
   \includegraphics[width=0.36\textwidth]{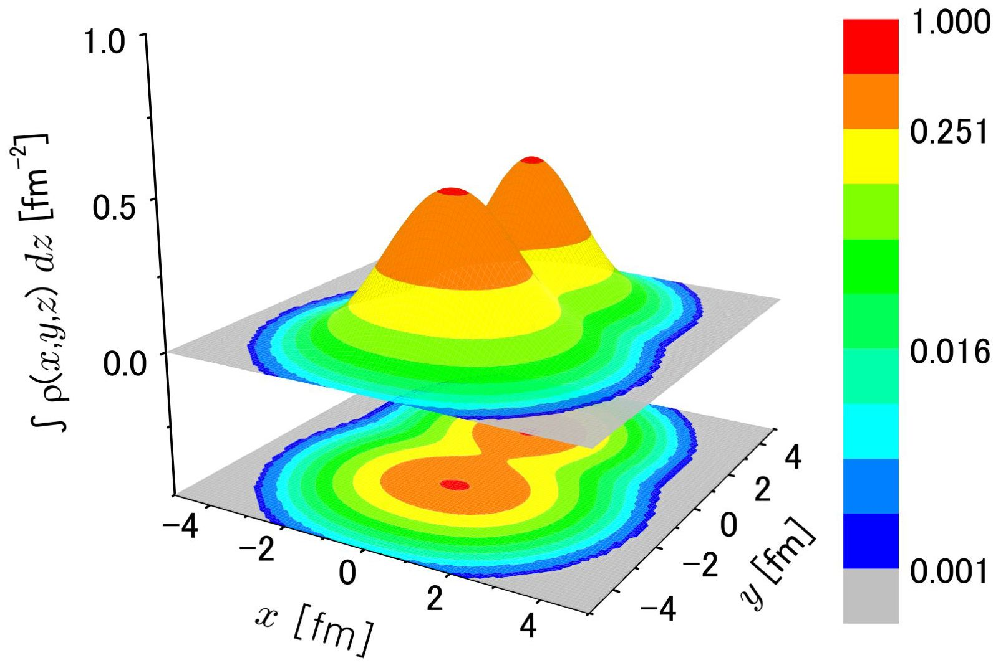}
  $ \ \ \ \ \ \ \ \ \ \ \ $
\label{fig:b1-motivation}
\end{figure}
\vspace{-0.1cm}
\noindent
{\footnotesize 
 {\bf FIGURE 3.} Density distribution of $^4$He
                 \cite{hksw-cluster}.
 \hspace{0.6cm}
 {\bf FIGURE 4.} Density distribution of $^9$Be by AMD
                 \cite{hksw-cluster}.
 }
\vspace{0.5cm}

\begin{wrapfigure}{r}{0.48\textwidth}
   \vspace{-0.3cm}
   \begin{center}
       \epsfig{file=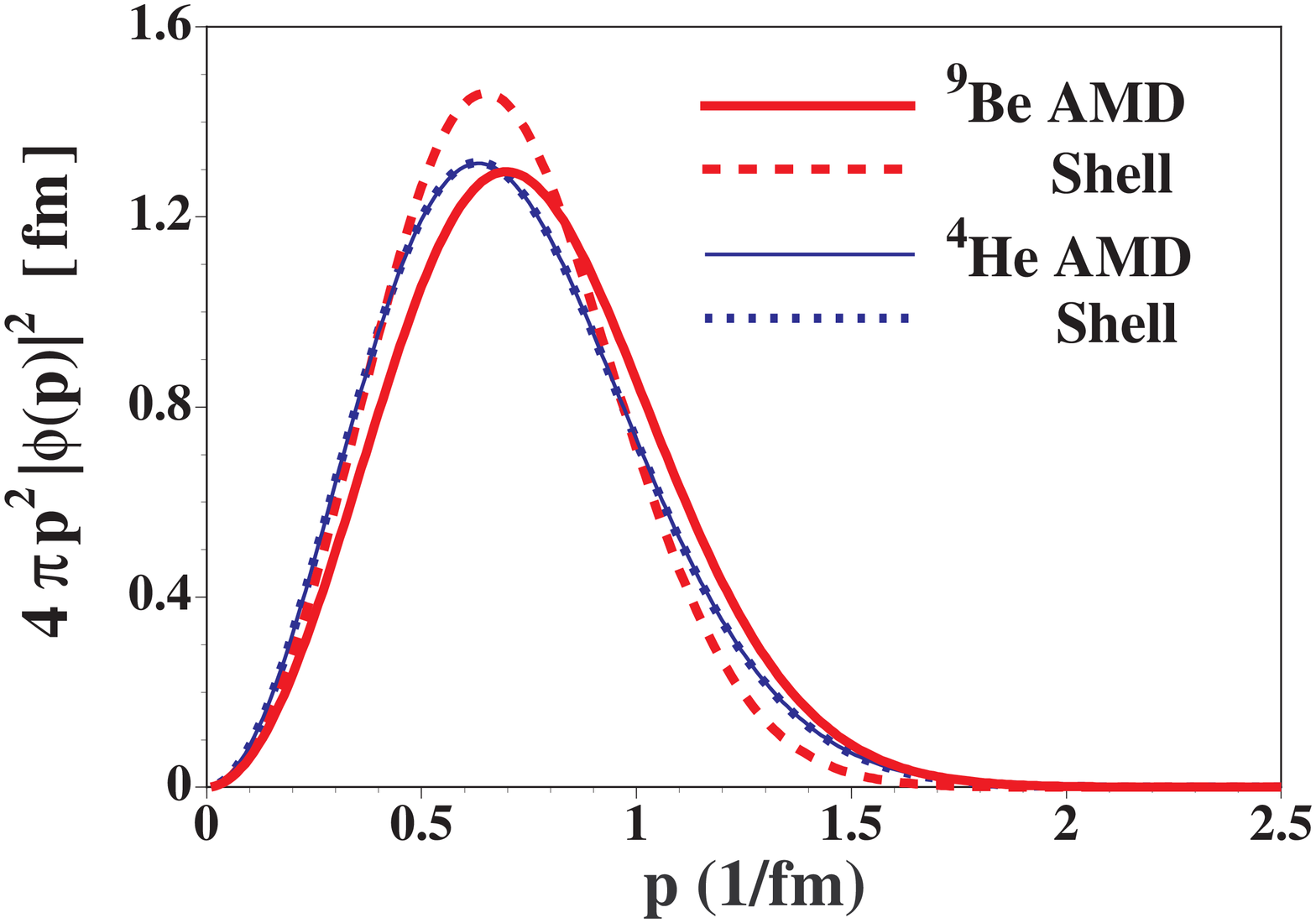,width=0.34\textwidth} \\
   \end{center}
   \vspace{-0.3cm}
{\footnotesize \hspace{0.2cm}
 {\bf FIGURE 5.} Nucleon-momentum distributions 

\noindent \hspace{0.24cm}
in $^4$He and $^9$Be by shell and AMD models \cite{hksw-cluster}.
 }
\vspace{0.4cm}
   \begin{center}
       \epsfig{file=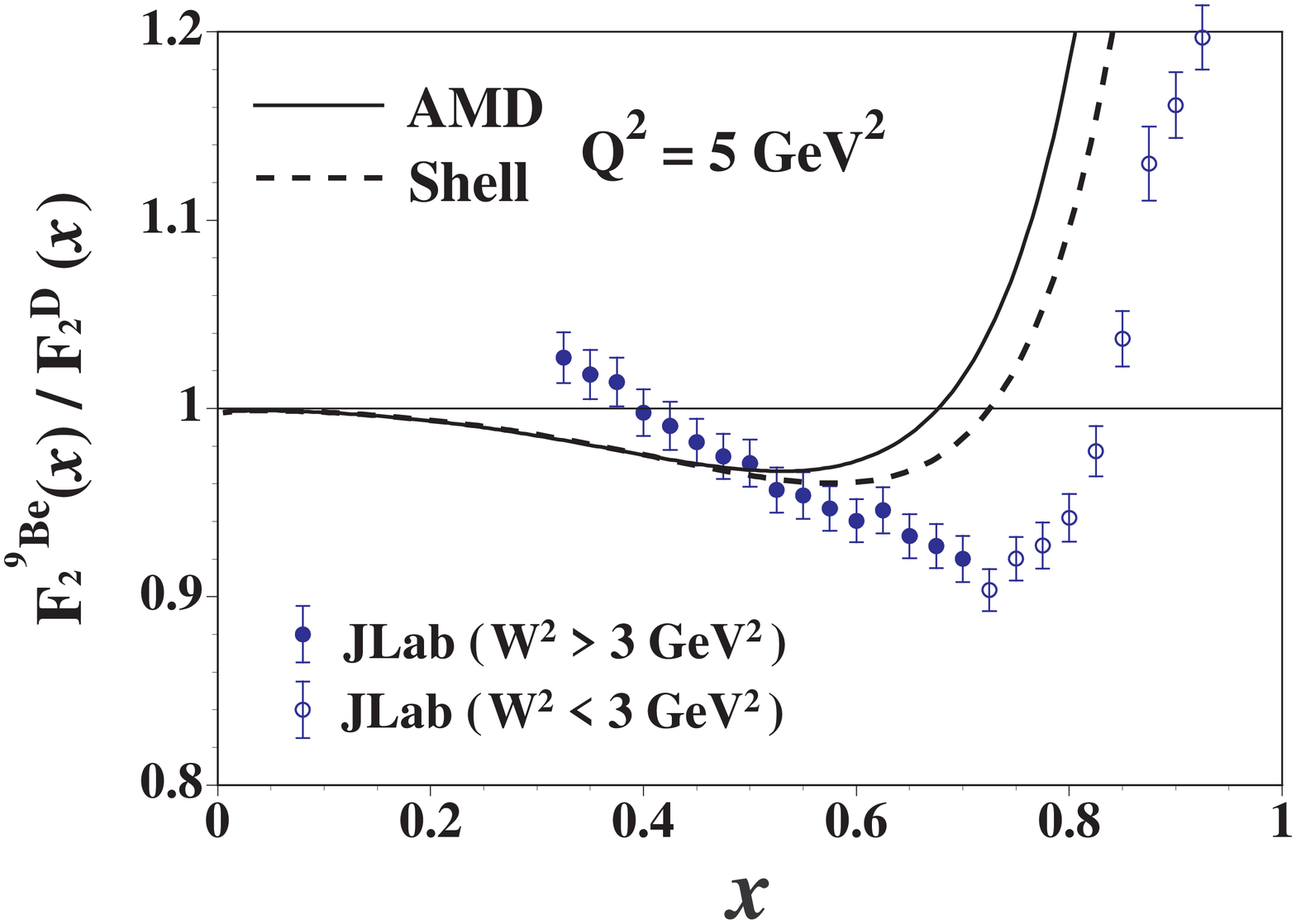,width=0.34\textwidth} \\
   \end{center}
   \vspace{-0.3cm}
{\footnotesize \hspace{0.75cm}
 {\bf FIGURE 6.} Nuclear modifications in $^9$Be
 
\noindent
\hspace{0.75cm}
                 by shell and AMD models \cite{hksw-cluster}.
 }
\vspace{-0.2cm}
\end{wrapfigure}

Although the two-dimensional ($x$, $y$) density for $^9$Be
in Fig. 4 is very different from a shell-model one, 
inhomogeneity of the nuclear density is washed out
if the angular average is taken \cite{hksw-cluster}. 
Then, transforming the densities to the momentum space, 
we obtain the results in Fig. 5 for $^4$He and $^9$Be. 
It is interesting to find that the AMD wave function of 
$^9$Be has larger high-momentum components than the shell-model
one, whereas both densities are the same in $^4$He.
The high-momentum components are created by the cluster
development in $^9$Be because nucleons are confined mainly
in the two small-space clusters. 

Using the momentum densities, we calculate the nuclear 
structure functions $F_2^A$ at $Q^2$=5 GeV$^2$
by using Eq.(\ref{eqn:w-convolution}).
The results are shown in Fig. 6 for $F_2^{^9 Be}/F_2^D$ 
together with experimental data.
Since the short-range correlations and internal
nucleon modifications are not taken into account
in our formalism, the curves do not fully agree with
the data. However, it is interesting to find some clustering
effects in $^9$Be as shown in Fig. 6 because there are
differences between the two curves of the shell and AMD models,
whereas both are the same in $^4$He.
If the current experimental errors and other effects
such as the correlations are considered, it may not be
easy to find a clear clustering signature in 
$F_2^A/F_2^D$ from the conventional mean part.

\begin{wrapfigure}{r}{0.43\textwidth}
   \vspace{-0.5cm}
   \begin{center}
       \epsfig{file=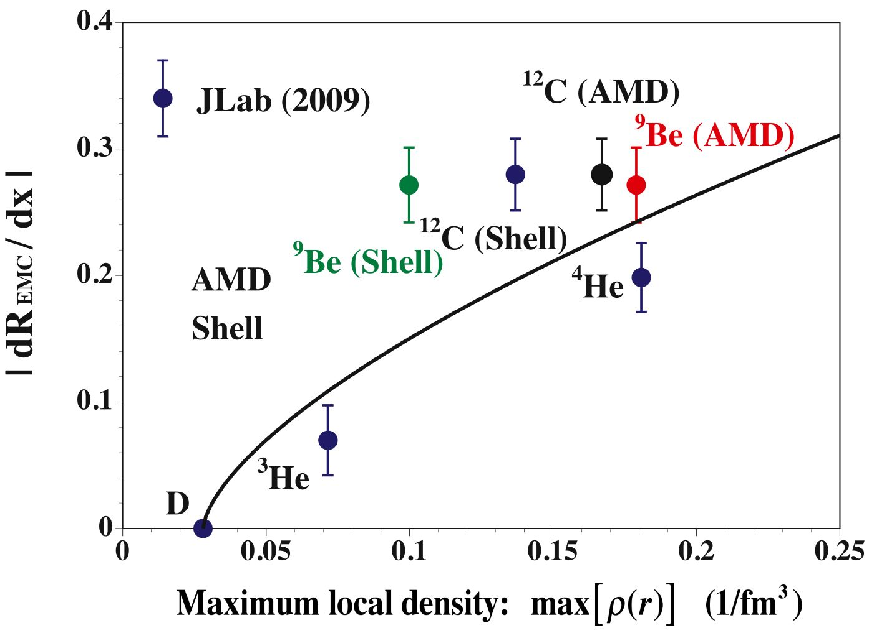,width=0.37\textwidth} \\
   \end{center}
   \vspace{-0.2cm}
{\footnotesize \hspace{0.30cm}
 {\bf FIGURE 7.} Nuclear modification slopes
 
\noindent
\hspace{0.30cm}
     shown by maximum local densities \cite{hksw-cluster}.
 }
\vspace{-0.0cm}
\end{wrapfigure}

We found that the conventional binding and Fermi-motion
contributions do not show a significant clustering effect.
Next, we consider the possibility that the high densities 
due to the cluster formation in $^9$Be could be the origin 
for the anomalous JLab data.
We plotted the JLab data by the maximum local densities
calculated by both models in Fig. 7. The curve interpolates
the shell-model data points except for $^9$Be. 
If the $^9$Be slope data is shown by 
the maximum density of the shell model, it is again 
too large to be expected from other nuclei. 
However, if it is shown by the maximum density of 
the AMD with the cluster structure, it is on the curve. 
In this way, the ``anomalous'' JLab result could be 
explained by the cluster structure in the nucleus.
There is also a small difference between 
the AMD and shell model in $^{12}$C of Fig. 7, which
is caused by the mixing of a cluster-like configuration 
in the AMD model of $^{12}$C.

Since it may be confusing for the reader that the $^9$Be slope 
is understood by the cluster structure although the effect
is rather small in Fig. 6, we would like to explain our viewpoint. 
The nuclear structure functions consist of the mean conventional 
part and the remaining one depending on the maximum local density:
\vspace{-0.05cm}
\begin{equation}
F_2^A = \text{(mean part)} 
       +\text{(part created by large densities
               due to cluster formation)}.
\end{equation}
\vspace{-0.05cm}
The first part is described by the usual convolution calculation
with the spectral function given by the averaged nuclear density 
distribution. The remaining part is associated with 
the inhomogeneity of the nuclear density, before taking 
the average of nuclear wave function, given by the nuclear cluster
structure. The physics behind the latter part could be
a nuclear-medium modification of internal nucleon structure.
Our studies suggest that the physics mechanism, associated 
with the high densities created by the clusters in $^9$Be,
could be the origin for the anomalous slope
$d(F_2^{^9 Be}/F_2^D)/dx$. Such studies of possible cluster structure 
in deep inelastic scattering will be continued at JLab 
\cite{jlab-cluster-exp}, and a more elaborated theoretical model 
needs to be developed for comparison with future data.

\vspace{-0.2cm}


\end{document}